\documentclass[aps,pra,twocolumn,groupedaddress,showpacs]{revtex4-1}

\usepackage{hyperref} 
\usepackage{bm} 
\usepackage{amssymb}
\usepackage{amsfonts}
\usepackage{amsmath}
\usepackage{slashed}
\usepackage[utf8]{inputenc}
\usepackage{graphicx}
\usepackage[mathscr]{euscript}

\newcommand{\fsp}[2]{{#1} \cdot {#2}}

\newcommand{\Nosc}{N_{\text{osc}}}
\newcommand{\pv}{\phi}	
\newcommand{\cep}{\chi}
\newcommand{\xim}{\xi_{\operatorname{max}}}

\newcommand{\vp}{{\bf p}}

\newcommand{\vk}{{\bf k}}

\newcommand{\eL}{\epsilon}
\newcommand{\veL}{{\bf \eL}}

\newcommand{\sk}{\slashed{k}}
\newcommand{\sA}{\slashed{A}}

\newcommand{\seL}{\slashed{\eL}}
\newcommand{\eg}{\epsilon_\gamma}
\newcommand{\seg}{\slashed{\epsilon}_\gamma}

\newcommand{\SMPuls}{\mathcal{S}}

\newcommand{\vn}{{\bf n}}
\newcommand{\vx}{{\bf x}}

\newcommand{\mn}{\bar{n}}
\newcommand{\amn}{|\mn|}
\newcommand{\mnp}{\bar{n}^\prime}
\newcommand{\amnp}{|\mnp|}

\newcommand{\Nmodes}{M}
\newcommand{\Nphotons}{N}

\newcommand{\Ppp}{p_{\mn}}
\newcommand{\PppN}{p_{\Nphotons}}
\newcommand{\Pph}{\varrho_{\mn}}
\newcommand{\PphN}{\varrho_{\Nphotons}}

\newcommand{\ft}{f(t)}
\newcommand{\fw}{\hat{f}(\omega)}

\newcommand{\pd}{\varrho} 

\newcommand{\cf}{\mathscr{X}} 

\begin{document}

\preprint{Ver. 1.0}

\title{Strong-Field Breit-Wheeler Pair Production in Short Laser Pulses:\\Identifying Multiphoton Interference and Carrier-Envelope Phase Effects}


\author{Martin J. A. Jansen}
\author{Carsten M\"uller}
\affiliation{Institut f\"ur Theoretische Physik I, Heinrich-Heine-Universit\"at D\"usseldorf, Universit\"atsstr. 1, 40225 D\"usseldorf, Germany}


\date{\today}

\begin{abstract}
The creation of electron-positron pairs by the strong-field Breit-Wheeler process in intense short laser pulses is investigated in the framework of laser-dressed quantum electrodynamics. 
Regarding laser field parameters in the multiphoton regime,
special attention is brought to the energy spectrum of the created particles, which can be reproduced and explained by means of an intuitive model. The model is based on the probabilities of multiphoton events driven by the spectral components of the laser pulse. It allows, in particular, to identify interferences between different pair production channels which exhibit a characteristic dependence on the laser carrier-envelope phase.
\end{abstract}



\maketitle

\section{Introduction}

The collision of two high-energy photons can lead to the creation of an electron-positron ($e^+e^-$) pair and is referred to as Breit-Wheeler process \cite{Breit1934}. Studies of Breit-Wheeler pair production via multiphoton absorption in the strong electromagnetic fields of intense laser waves started in the 1960s \cite{Reiss1962,Nikishov1964,Narozhny1965}. Here, the energy threshold for pair production is overcome by the combination of a high-energy $\gamma$ photon and several laser photons, according to the reaction
\begin{eqnarray}
\label{BW}
\omega_\gamma + n \omega_c \to e^+e^-\ .
\end{eqnarray}
Due to a remarkable and still ongoing progress in high-power laser technology, there is a clear perspective for corresponding experimental studies in the near future. Petawatt-class laser systems reach field intensities well above 10$^{20}$\,W/cm$^2$ in many laboratories worldwide, and an increase towards 10$^{25}$\,W/cm$^2$ is envisaged within the Extreme Light Infrastructure (ELI) \cite{ELI} and the Exawatt Center for Extreme Light Studies (XCELS) \cite{XCELS}. Besides, x-ray free-electron laser facilities can generate brilliant kiloelectronvolt photon beams beyond 10$^{20}$\,W/cm$^2$ \cite{DESY,LCLS}. These technological developments have triggered substantial theoretical efforts on $e^+e^-$ pair production and other quantum electrodynamic processes in high-intensity laser fields during the last decade \cite{Ehlotzky_Review,DiPiazza2012}.

Operating modern technologies in a suitable conjunction allows for experimental studies on $e^+e^-$ pair production in strong laser fields already at present. Indeed, the first (and so far unique) experimental observation of multiphoton Breit-Wheeler pair production was accomplished in the 1990s by utilizing  ultrarelativistic electron-laser collisions at the Stanford Linear Accelerator Center (SLAC) \cite{Burke1997}. In the rest frame of the projectile electrons, the 
laser field strength and frequency were largely amplified to the required level by the relativistic Lorentz boost. The resulting pair creation was attributed to processes involving on average four to five laser photons and a high-energy $\gamma$ photon which originated from Compton backscattering off the electron beam. We note that $e^+e^-$ pairs can also be generated when intense laser pulses interact with solid targets \cite{Chen2009,Sarri2013}. Here, however, the pair production is not caused by multiphoton absorption, but rather by a chain of reactions involving emission of bremsstrahlung and (Bethe-Heitler) pair production in the Coulomb fields of atomic nuclei in the target.

While the seminal works \cite{Reiss1962,Nikishov1964,Narozhny1965} were based on the assumption of an infinitely extended, monochromatic laser wave, in recent years theoreticians have considered refined aspects of laser-induced pair production in more complex field geometries. Two main lines of development may be distinguished. On the one hand, processes in bichromatic laser fields have been investigated \cite{Narozhny2000,Krajewska2012BH_Phase,Augustin2013,Xue2014,Jansen2015}. These fields facilitate interferences between pair creation channels of different photon number combinations, which can be modulated by the relative phase shift between the laser modes. Furthermore, suitable combinations of a strong, low-frequency and a weak, high-frequency field mode were shown to enhance the pair creation yield significantly \cite{Schuetzhold2008,DiPiazza2009,Orthaber2011,Grobe1,Jansen2013,Akal2014,Otto2015}.
On the other hand, in view of the experimental prospects for $e^+e^-$ pair production in strong laser fields, theoretical treatments have started to account for the actual shape of finite laser pulses \cite{Fedorov2006,Heinzl2010,Ipp2011,Titov2012,Roshchupkin2012,Nousch2012,Krajewska2012BW,Grobe2,Krajewska2013BH_Pulse,Fedotov2013,Hu2014,Krajewska2014,Meuren2014,Meuren2015,Nousch2015,Titov2015}. 
The broad spectrum of a short pulse was shown to facilitate a multitude of pair creation channels and to strongly modify the created particle spectra. The latter exhibit a quite complicated structure which can also be affected by the carrier-envelope phase (CEP) of the pulse. In addition, characteristic enhancements of the pair production probability were found in certain laser parameter regimes. 

In this paper, $e^+e^-$ pair creation by the multiphoton Breit-Wheeler process \eqref{BW} in an intense laser pulse is examined. Laser fields with a moderate value of the normalized vector potential, $\xi\le1$, will be considered [see Eq.\,\eqref{xi} below]. We shall combine concepts from both theoretical approaches mentioned above in order to show how the energy spectrum of the created particles can be understood in terms of multiphoton processes originating from the continuous pulse spectrum. The origins of CEP effects in the particle spectra shall be classified and essentially traced back to interferences between these channels, which this way can be identified and understood. 
Our analysis is based on an intuitive model which accounts for the probability of multiphoton events driven by the laser pulse. 
It also allows us to identify combined emission-absorption processes of laser photons in the particle spectra. We note that another simplified model of multiphoton Breit-Wheeler pair production has been developed in \cite{Nousch2012} which, however, conceptually differs from our approach, as will be discussed below.
The present study, moreover, complements a recent article \cite{Meuren2015}, where interference processes in a laser field with $\xi\gg1$ are interpreted on the basis of a semiclassical model.

We point out that the Breit-Wheeler process under consideration here is related, by a crossing symmetry of the underlying Furry-Feynman diagrams, to nonlinear Compton scattering. For the latter process, signatures of short laser pulses -- including interference and CEP effects -- have also been studied in recent years (see, e.g., \cite{Boca2009,Mackenroth2010,Heinzl2010C,Krajewska2012Co,Harvey2015}).
Besides, it is worth mentioning that interesting analogies of strong-field $e^+e^-$ pair production exist in other areas of physics. They have been theoretically predicted in various systems, such as graphene layers in external electric fields \cite{Allor2008,Klimchitskaya2013} or ultracold atoms in optical lattices \cite{Szpak2011,*Szpak2012}, where the dynamics of quasiparticle and hole states can resemble the $e^+e^-$ pair creation.

Our paper is organized as follows:
We start with a brief survey of the theoretical framework in Sec. \ref{Sec:Theo} and present the analytical derivation of the pair creation probability. In Sec. \ref{Sec:Mod}, concepts from a bichromatic laser field are generalized, revealing different types of interferences that are facilitated by the continuous spectrum of short laser pulses. A model based on multiphoton processes is developed and shown to reproduce the particle energy spectra to a very good approximation. In Sec. \ref{Sec:Int}, CEP effects are investigated and classified. Employing information obtained from the model, interferences between different pair production channels are identified. 
We finish with concluding remarks in Sec. \ref{Sec:Conclusion}.

Throughout this work, relativistic units with $\hbar=c=1$ are used. Furthermore, $e$ and $m$ denote the positron charge and mass, respectively.

\section{Theoretical Framework}
\label{Sec:Theo}
The strong-field Breit-Wheeler process describes the creation of an $e^+e^-$ pair induced by the decay of a high-energy gamma quantum traveling in a strong laser field. 
The corresponding pair creation probability is obtained following an $S$ matrix approach in the framework of laser-dressed quantum electrodynamics (QED).
We will settle our notation and definitions in Sec. \ref{Sec:TF_1} and present a brief sketch of the calculation in Sec. \ref{Sec:TF_2}.
A more detailed presentation of a similar calculation can be found in \cite{Krajewska2012BW}.
\subsection{Definitions}\label{Sec:TF_1}
\subsubsection{Laser Pulse}

The laser pulse is described classically by its vector potential in radiation gauge
\begin{equation}\label{VecPot}
 A^\mu = A_0 f(\pv) \cf_{[0,2\pi]}(\pv) \eL^\mu \,.
\end{equation}
The space-time dependence is determined by the phase variable $\pv = \fsp{k}{x}$, leading to a plane-wave fronted pulse and facilitating the use of Volkov states. 
We introduce a wave four-vector $k^\mu=\omega_b(1,{\bf n})$ with (basic) frequency $\omega_b$ and normalized propagation direction ${\bf n}$, a real polarization vector $\veL^\mu$ with $\fsp{\eL}{k}=0$ and the  space-time coordinate $x^\mu=(t,{\bf r})$.
The actual shape is given by the combination of the shape function $f(\pv)$ and the characteristic function $\cf_{[0,2\pi]}(\pv)$, which restricts the pulse to the phase interval $[0,2\pi]$. 
Even though the following derivation is kept general, we will present our numerical results for a specific choice of the shape function which is defined by means of its derivative
\begin{equation}\label{our_pulse}
 f^\prime(\pv)=\sin^2(\pv/2)\sin(\Nosc\pv+\cep)\,,
\end{equation}
with the number $\Nosc$ of cycles within the $\sin^2$-envelope, and a carrier-envelope phase $\cep$. 
The spectrum of this pulse is centered around the central frequency $\omega_c=\Nosc\omega_b$.
In order to fulfill Maxwell's equations, we have to restrict $\Nosc\geq2$; otherwise the vector potential in Eq. \eqref{VecPot} would not be continuous.

The strength of the laser field is measured with the invariant and dimensionless amplitude parameter
\begin{equation}
\label{xi}
 \xim = \frac{eA_0}{m} \operatorname{max}_\pv |f(\pv)|
\end{equation}
accounting for the maximum amplitude, which is most important for the non-linear processes that we are interested in. 

\subsubsection{Volkov States}
The Dirac equation for a charged particle in the above electromagnetic field can be solved exactly by means of Volkov states due to the restriction of the space-time dependence on one common propagation direction \cite{LandauLifshitz}.
For a particle (antiparticle) with four-momentum $p_-^\mu$ ($p_+^\mu$), the Volkov states read
\begin{equation}
 \Phi_{p_\pm} = \sqrt{\frac{m}{V E_{p_\pm}}} \left[ 1 \pm \frac{e\sk\sA}{2 \fsp{k}{p_\pm}} \right] w_{\pm} e^{i[\pm \fsp{p_\pm}{x}+\Lambda_\pm]}
\end{equation}
with
\begin{equation}
 \Lambda_\pm = \frac{1}{\fsp{k}{p_\pm}} \int_0^{\fsp{k}{x}} \left[ e \fsp{p_\pm}{A(\pv)} \mp \frac{e^2}{2} A^2(\pv) \right] d\pv
\end{equation}
and with $p_\pm^\mu = (E_{p_\pm},\vp_\pm)$, energy $E_{p_\pm}=\sqrt{m^2+\vp_\pm^2}$ and a normalizing volume $V$. 
The free spinors $w_{\pm}$ satisfy the algebraic equation $(\slashed{p} \pm m) w_{\pm} = 0$ and are normalized according to $\overline{w}_{\pm}(p_\pm,s_\pm) w_{\pm}(p_\pm,s^\prime_\pm) = \mp \delta_{s_\pm,s^\prime_\pm}$,
where $s_\pm$ labels the spin projection \cite{GreinerRelQM}. Since we will sum over all spin configurations in the end, the choice of the quantization axis is immaterial for this consideration.
We employ the metric tensor $\operatorname{diag}(+,-,-,-)$, Dirac $\gamma$-matrices, the Dirac adjoint $\overline{w}=w^\dagger \gamma_0$, and Feynman slash notation.

The interaction with the laser field gives rise to effective momenta
\begin{equation}\label{EffMomentum}
 q_\pm^\mu = p_\pm^\mu \mp e \langle A^\mu \rangle \pm  \langle  e \fsp{p_\pm}{A} \mp \frac{e^2}{2} A^2 \rangle  \frac{k^\mu}{\fsp{k}{p_\pm}}\,,
\end{equation}
where $\langle \cdots \rangle$ indicates a phase average. 
Note that, in contrast to an infinitely extended monochromatic wave, the average over $A^\mu$ does not need to be zero.
Signatures of these dressing effects can be detected in the energy spectra of produced particles.

\subsubsection{Gamma Quantum}
The gamma quantum is treated as one mode of a quantized radiation field with wave four-vector $k_\gamma^\mu = (\omega_\gamma,{\bf k}_\gamma)$, real polarization vector $\eg^\mu$ with $\fsp{k_\gamma}{\eg}=0$ and corresponding mode index $\lambda_\gamma$. 
For calculational simplicity, we assume the gamma quantum to be colliding head-on with the laser beam.

The absorption of the gamma quantum is included into the effective scattering potential
\begin{equation}\label{GammaQuantum}
  A_\gamma^\mu = \sqrt{\frac{2\pi}{V\omega_\gamma}} e^{-i\fsp{k_\gamma}{ x}} \eg^\mu 
\end{equation}
which facilitates concise notation of the following expressions.

\subsection{Pair Creation Probability}\label{Sec:TF_2}
The $S$ matrix element for the gamma photon-induced creation of an electron-positron pair with four-momenta $p_-$  and $p_+$ reads
\begin{equation}
 \SMPuls_{p_+p_-} = ie \int d^4x \, \overline{\Phi}_{p_-} \slashed{A}_\gamma \Phi_{p_+}\,.
\end{equation}

Sorting the constituents with respect to their dependence on the integration variables,
we arrive at
\begin{equation}\label{SMatrix_Compact}
  \SMPuls_{p_+p_-} = S_0 \int d^4x \, C(\pv) e^{-i\fsp{Q}{x}-iH(\pv)}\,,
\end{equation}
with $S_0 = iem \sqrt{ 2\pi / ( V^3 \omega_\gamma E_{p_+} E_{p_-})} $ and

\begin{eqnarray}
 C(\pv) &=& g_0 + g_1 f(\pv) \cf_{[0,2\pi]}(\pv) \nonumber \\
 H(\pv) &=& \int_0^\pv h(\tilde{\pv}) d\tilde{\pv} \\
 Q^\mu &=& k_\gamma^\mu - (p_+^\mu + p_-^\mu) \nonumber
\end{eqnarray}
and introducing abbreviations
\begin{eqnarray}
 g_0 &=& \overline{w}_- \seg w_+ \nonumber \\
 g_1 &=& \frac{eA_0}{2} \overline{w}_- \left[ \frac{\seg\sk\seL}{\fsp{k}{p_+}}  - \frac{\seL\sk\seg}{\fsp{k}{p_-}}  \right] w_+ \nonumber \\
 h(\pv) &=& \left[ h_1 f(\pv)+h_2 f(\pv)^2\right] \cf_{[0,2\pi]}(\pv) \\
 h_1 &=&  - eA_0 \left[ \frac{\fsp{\eL}{p_+}}{\fsp{k}{p_+}} - \frac{\fsp{\eL}{p_-}}{\fsp{k}{p_-}} \right] \nonumber \\
 h_2 &=& - \frac{e^2A_0^2}{2} \left[ \frac{1}{\fsp{k}{p_+}} + \frac{1}{\fsp{k}{p_-}} \right] \nonumber \,.
\end{eqnarray}

We employ light-cone coordinates with respect to the laser propagation direction $\vn = \vk / |\vk|$. For a given four-vector $x^\mu$ with $x^\parallel = \vx \cdot \vn$ we have $x^- = x^0-x^\parallel$, $x^+=\frac{1}{2}\left(x^0+x^\parallel\right)$ and $\vx^\bot = \vx - x^\parallel \vn$. The integration measure is transformed according to $d^4 x = dx^-dx^+d^2 x^\bot$. 

The integration in Eq. \eqref{SMatrix_Compact} requires a regularization, which was done in analogy to the Boca-Florescu transformation \cite{Boca2009}.
Effectively, we have to set
\begin{equation}
 g_0 \rightarrow \frac{-k^0}{Q^0} h(\pv) g_0
\end{equation}
where $Q^0\neq 0$ follows from kinematical constraints \cite{Krajewska2012BW}.

Three integrations can be carried out directly, and we obtain
\begin{eqnarray}\label{SMatrix_Final}
 \SMPuls_{p_+p_-} &=& (2\pi)^3 S_0 \delta(Q^-) \delta^{(2)} ({\bf Q}^\bot) \\*
&\times& \int_0^{2\pi/k^0} dx^- C(k^0x^-) e^{-iQ^0x^--iH(k^0x^-)} \nonumber \,.
\end{eqnarray}
The last integration can be carried out numerically.

The total particle creation probability (averaged over the polarization of the gamma quantum) is given by
\begin{equation}\label{P_Pulse}
 \mathscr{P} = \frac{1}{2} \sum_{\lambda_\gamma} \sum_{s_+,s_-} \int \frac{V d^3 p_+}{(2\pi)^3} \int \frac{V d^3 p_-}{(2\pi)^3} |\SMPuls_{p_+p_-}|^2 \,.
\end{equation}
The spinor properties can be treated with usual trace techniques. 
We note that the square of $2\pi\delta(Q^-)$, which appears in Eq. \eqref{SMatrix_Final}, has to be treated with care. It produces a scaled length factor of $\frac{k_\gamma^0}{k_\gamma^-}\,L$, instead of just $L$. This can be shown, for example, within a consideration which treats the incoming $\gamma$-photon as a wave packet \cite{Meuren2014}. Here, $L$ denotes the normalizing length in the propagation direction [cp. Eq. \eqref{GammaQuantum}]. In our geometry of a head-on collision, the scaling factor simply becomes $\frac{k_\gamma^0}{k_\gamma^-} = \frac{1}{2}$. It enters into the particle creation probability Eq.  \eqref{P_Pulse} through the square of the $S$ matrix element.

\subsection{Energy-Momentum Balance}
The three $\delta$-functions in Eq. \eqref{SMatrix_Final} can be used to formulate a pair formation condition in terms of the dressed particle momenta
\begin{equation}\label{mom_bal_dre}
 q_+^\mu + q_-^\mu = k_\gamma^\mu + r k^\mu \,.
\end{equation}
We note that a consideration of the particles' dressed momenta appears physically meaningful in the present situation because, for $\xi<1$, the pair formation length covers a substantial fraction of the laser pulse \cite{DiPiazza2012}, which comprises several cycles of field oscillations.
Equation \eqref{mom_bal_dre} closely resembles the energy-momentum conservation law in an infinitely extended, monochromatic laser field \cite{Reiss1962,Nikishov1964,Narozhny1965}, with the only difference being that the parameter $r$ is continuous, rather than discrete. Accordingly, $rk^\mu$ measures the four-momentum absorbed from the laser pulse \cite{Krajewska2012BW}.
As a consequence, the parameter $r$ becomes subject to the threshold condition $r\ge\frac{m_*^2}{\omega_b \omega_\gamma}$ which is formulated in terms of the laser-dressed mass 
\begin{equation}
m_* = \sqrt{q_\pm^2} = \sqrt{ m^2 + e^2 \langle A \rangle^2 - e^2 \langle A^2 \rangle}\,. 
\end{equation}
Since $r$ is not restricted to integer values, each frequency component of the pulse spectrum can in principle participate in the pair production process, given that the threshold is overcome.
Regarding multi\-photon processes, Eq. \eqref{mom_bal_dre} does not specify how the total four-momentum $rk^\mu$ is shared among several laser photons. 
In fact, the production of a pair with fixed four-momenta can be thought of as a coherent superposition of channels of different numbers of photons with continuously varying frequencies as provided by the pulse spectrum.
One aim of this article is to identify the dominant contribution to such a process.

As a preparation for our further analysis of the particle energy spectra, we introduce $E_L = rk^0$ to denote the total required energy from the laser, with
\begin{equation}\label{dressing}
 rk^\mu = p_+^\mu + p_-^\mu -k_\gamma^\mu - \left( h_1 \langle f \rangle  + h_2 \langle f^2 \rangle\right) k^\mu \,.
\end{equation}

\section{Modeling the Particle Spectrum}
\label{Sec:Mod}
The aim of this section is to develop an intuitive model for the pair creation process in a laser pulse which accounts for effects due the pulse spectrum. As a first step, we generalize concepts from a bichromatic to a multichromatic field. Afterwards, these concepts will be applied to an actual laser pulse which can be considered the continuous limit.

\subsection{Concepts from Multichromatic Case}\label{Sec:Multichrom_Field}
\subsubsection{Pair Creation Amplitude}
For the case of a multichromatic laser field with $\Nmodes$ discrete modes of frequencies $\omega_j$ (which are assumed pairwise different) and relative phase shifts $\delta_j$ propagating in the same direction, the combined vector potential in reduced units reads
\begin{equation}
\label{Multichrom_Field}
 \xi(t) = \sum_{j=1}^\Nmodes \xi_j \cos( \omega_j t - \delta_j)
\end{equation}
where the usual amplitude parameter $\xi_j = \frac{eA_{0,j}}{m}$ has been used.
We suppress the spatial dependence in the notation. The polarization vectors are assumed to be linear and to be aligned uniformly, even though we do not explicitly use the latter property \footnote{Compare the calculation in \cite{Narozhny2000} which holds for arbitrary angles between the polarization vectors.}.

The pair creation amplitude $\mathcal{S}$ for this laser field can be Fourier decomposed into the form
\begin{equation}\label{S_Multi}
 \mathcal{S} = \sum_{n_1=-\infty}^\infty \dots \sum_{n_\Nmodes=-\infty}^\infty \mathcal{S}(n_1,\dots,n_\Nmodes)
\end{equation}
with partial amplitudes 
\begin{equation}\label{S_Multi_part} 
\mathcal{S}(\mn) = \mathcal{S}_0(\mn) \, \delta( q_+^0 + q_-^0 -\omega_\gamma - \sum_{j=1}^\Nmodes n_j \omega_j) \, e^{i \varphi(\mn)}
\end{equation}
that are referenced via the multiindex $\mn=(n_1,\dots,n_\Nmodes)$. Inspecting the energy conservation for the pair creation process, we identify the energy
$q_+^0 + q_-^0 -\omega_\gamma$
which is required from the laser field in order to create a certain particle pair.
The $\delta$-function allows only those processes, for which this energy can be met by integer multiples of the mode frequencies $\omega_j$.
This step finally brings the concept of (laser) photons into our framework despite the classical treatment of the laser field.

Note that the numbers $n_j$ in Eq. \eqref{S_Multi} can also be negative, which means that the corresponding photons are emitted into the laser field \cite{Xue2014,Akal2014}.
Depending on the availability of other photons, these processes can deliver the dominant contribution for certain particle energies.

The phase of the pair creation amplitude
\begin{equation}
\label{AmplitudePhase}
 \varphi(\mn) = \sum_{j=1}^\Nmodes n_j \delta_j
\end{equation}
results as the sum of the phase shifts of the contributing modes.

In order to classify the subprocesses in the following section, we introduce the symbol $\amn = \sum_{j=1}^\Nmodes |n_j|$, and rearrange the amplitude as 
\begin{equation}
 \mathcal{S} = \sum_{\mn} \mathcal{S}(\mn) = \sum_{\Nphotons=1}^\infty \sum_{\amn=\Nphotons} \mathcal{S}(\mn) = \sum_{\Nphotons=1}^\infty \mathcal{S}(\Nphotons)
\end{equation}
where the $\Nphotons$-photon amplitude 
$\mathcal{S}(\Nphotons)=\sum_{\amn=\Nphotons} \mathcal{S}(\mn) $
has been introduced, such that $\Nphotons$ accounts for the {\it total} number of photons being interchanged with the laser field.

\subsubsection{Probabilities and Interferences}
The pair creation probability $\mathscr{P}$ is obtained by integrating $|\mathcal{S}|^2$ over the particle momenta.
We divide the absolute square of the amplitude into equal and non-equal total photon-number contributions
\begin{eqnarray}
 |\mathcal{S}|^2 = \sum_\Nphotons \bigg[ \mathcal{S}(\Nphotons) \mathcal{S}^* (\Nphotons) + \sum_{\Nphotons^\prime \neq \Nphotons} \mathcal{S}(\Nphotons) \mathcal{S}^* (\Nphotons^\prime) \bigg] 
\end{eqnarray}
and introduce the fully differential probabilities
$ \mathcal{P}_{\Nphotons} = \mathcal{S}(\Nphotons) \mathcal{S}^* (\Nphotons)$
and
$ \mathcal{P}_{\Nphotons\Nphotons^\prime} = \mathcal{S}(\Nphotons) \mathcal{S}^* (\Nphotons^\prime)+  \mathcal{S}(\Nphotons^\prime) \mathcal{S}^* (\Nphotons)$, where the latter results from interferences between channels of $\Nphotons$ and $\Nphotons^\prime \neq \Nphotons$ photons.
As we will see in the following, also $\mathcal{P}_\Nphotons$ can include interferences.

The criterion for the existence of interferences between two processes with photon combinations $\mn$ and $\mnp \neq \mn$ can be stated as \cite{Narozhny2000,Augustin2013,Xue2014,Jansen2015}
\begin{equation}\label{pnc}
 \sum_{j=1}^\Nmodes n_j \omega_j = \sum_{j=1}^\Nmodes n^\prime_j \omega_j \,.
\end{equation}
In the case of a bichromatic field, and when emission processes are negligble, this condition can only be fulfilled with $N^\prime \neq N$. In the general case, in particular when proceeding to a multichromatic field with $\Nmodes>2$, interferences can also occur between channels of the same total number of photons ($N^\prime = N)$, which we shall call Self-Interferences.

These processes become apparent, when we decompose
the $\Nphotons$-photon probability $\mathcal{P}_\Nphotons$ according to
\begin{equation}
 \mathcal{P}_\Nphotons =\sum_{\amn=\Nphotons} \bigg[ \mathcal{S}(\mn) \mathcal{S}^*(\mn) + \sum_{\substack{\amnp=\Nphotons \\ \mnp\neq\mn}}\mathcal{S}(\mn) \mathcal{S}^*(\mnp) \bigg] \,,
\end{equation}
where we recognize the ordinary $\Nphotons$-photon probability
$\mathcal{P}_\Nphotons^{\text{ord}} = \sum_{\amn=\Nphotons} \mathcal{S}(\mn) \mathcal{S}^*(\mn) $.
The remaining contribution to the probability of the $\Nphotons$-photon process is given by the ($\Nphotons$-photon) Self-Interference probability
\begin{equation}
 \mathcal{P}_\Nphotons^{\text{SI}} = \sum_{\amn=\Nphotons} \sum_{\substack{\amnp=\Nphotons \\ \mnp\neq\mn}}\mathcal{S}(\mn) \mathcal{S}^*(\mnp) \,,
\end{equation}
which consists of interferences between different combinations of the same total number $\Nphotons$ of photons. 

The interference terms between channels of $N$ and $N^\prime \neq N$ photons are of the form
\begin{equation}\label{IntTerms}
 \mathcal{P}_{N N^\prime} = 2\sum_{\substack{\amn=N \\  \amnp=N^\prime}}  |\mathcal{S}_0(\mn)| |\mathcal{S}_0(\mnp)| \sigma_{\mn,\mnp}
 \cos\left[{\varphi}(\mn)-{\varphi}(\mnp)\right]
\end{equation}
when the $\delta$-functions are suppressed, and assuming $\sigma_{\mn,\mnp} = \cos(\arg [\mathcal{S}_0(\mn)] - \arg [\mathcal{S}_0(\mnp)]) = \pm1$ .
The interference terms are modulated by the optical phase shifts $\delta_j$ [cp. Eq. \eqref{AmplitudePhase}], which can thus induce strong effects on the particle yield.

If the amplitude phases depend only on the total number of photons, allowing to write  $\varphi(\mn)=\varphi(N)$, we can define a common interference phase 
\begin{equation}\label{Int_N_Photon}
 \phi_{\Nphotons\Nphotons^\prime} = \varphi(N)-\varphi(N^\prime)\,.
\end{equation}
As we will show below, several of its properties can be derived analytically.

\subsection{P-Model}
Our aim is to understand the energy spectra of emitted positrons. To this end, we will now develop a model which produces quantitative estimates for the (ordinary) probabilities of processes depending on the total numbers of photons. 
Neglecting the self-interference terms, this approach facilitates a straight-forward implementation and additionally allows to estimate the magnitude and phase of interference terms.

\subsubsection{Definitions}\label{Sec:PM:Pre}
In order to employ the concepts from the previous section, we introduce for a given fully differential probability $\mathcal{P}$ the corresponding contribution to the total probability
\begin{equation}\label{IntoObs}
 \mathscr{P} = \frac{1}{2} \sum_{\lambda_\gamma} \sum_{s_+,s_-} \int \frac{V d^3 p_+}{(2\pi)^3} \int \frac{V d^3 p_-}{(2\pi)^3} \mathcal{P}
\end{equation}
which is supposed to be applied to the probabilities of partial processes, like $\mathcal{P}_N$ or $\mathcal{P}_{N N^\prime}$.

Our energy spectra are defined as follows:
The positron energy is scanned for a fixed positron emission direction.
The resulting data is then presented as a function of the required energy from the laser, where dressing effects are included  [cp. Eq. \eqref{dressing}].
We introduce the symbol $d\mathscr{P}$ to refer to the corresponding differential (spectral) probability 
\footnote{The definition in Eq. \eqref{dP} should not be confused with $\frac{d^3 \mathscr{P}}{d E_L d^2 \Omega_{e^+}}$ which includes up to two different positron energies for a given value of $E_L$. This is a purely kinematical effect which can be understood by regarding the creation of a pair in the forward and backward direction of the laser propagation axis in the c.m. system, where both particles have the same energy. Boosting this process along the laser axis, both particles may end up traveling along with the laser beam, but with different energies. Since the kinematics are invariant under the exchange of the particles, two energy solutions for positrons being emitted in the forward direction appear.},
which is given by
\begin{equation}\label{dP}
 d\mathscr{P} = \frac{d^3 \mathscr{P}}{d E_{p_+} d^2 \Omega_{e^+}} \frac{\partial E_{p_+}}{\partial E_L}\,.
\end{equation}
These energy spectra can be defined for both the full calculation from Sec. \ref{Sec:Theo} and for partial processes as introduced in Sec. \ref{Sec:Multichrom_Field}.

\subsubsection{Outline of the P-Model} 
The probability corresponding to $\mathcal{P}_{\mn}^{\text{ord}} = \mathcal{S}(\mn) \mathcal{S}^*(\mn)$ can be understood as the classical probability to create a pair from the photon combination $\mn$. Here, classical means the absence of interference processes.
The model idea is to decompose the probability to create a pair with total required energy $E_L$ into the probability $\Pph(E_L)$ to find a suitable photon combination $\mn$ in the frequency spectrum of the laser pulse, and the probability $p_{\mn}(E_L)$ to create a particle pair from these photons, 
which means
\begin{equation}
 d \mathscr{P}_{\mn}^{\text{ord}} (E_L) \approx p_{\mn}(E_L) \Pph(E_L) \,.
\end{equation}
In analogy to the $\delta$-function in Eq. \eqref{S_Multi_part}, $\Pph$ determines whether the laser field can provide the required four-momentum in form of the particular photon combination $\mn$ in order to produce a given particle pair. Besides, $\Pph$ accounts quantitatively for the spectral shape of the laser field.
Conversely, the field intensity determines the pair creation probability $p_{\mn}$.

While $\Pph$ will be determined exactly, we develop the model by simplifying the pair creation probability. 
As a first step, we assume that $\Ppp(E_L)$ does not depend on the distribution of the photons $\mn$, but only on their total number $\Nphotons=\amn$, which means $\Ppp(E_L)\approx \PppN(E_L)$.

Introducing the probability
$\PphN(E_L) = \sum_{\amn=\Nphotons} \Pph(E_L)$
to find any combination of $\Nphotons$ photons that delivers the required energy,
we arrive at the probability
\begin{equation}\label{Model_P_Ord}
 d\mathscr{P}_{\Nphotons}^{\text{ord}} (E_L) \approx \PppN(E_L) \PphN(E_L)
\end{equation}
of the (ordinary) pair creation channel which includes $N$ photons.
Effectively, we have separated the spectral properties of the laser pulse from the pair creation process.

The model approach is obtained by assuming the pair creation probabilities to be mainly determined by the perturbative intensity scaling
\begin{equation}\label{Model_Scaling}
 \PppN (E_L) = \mathsf{p}_0 \, \xim^{2\Nphotons}
\end{equation}
with $\xim$ as given by Eq. \eqref{xi}.
The remaining factor $\mathsf{p}_0$ can be regarded as a global prefactor for a given particle energy spetrum. As we will show below, the resulting estimate can nicely reproduce various features of the particle spectra.

Accordingly, interference terms can be modelled by
\begin{equation}
 d\mathscr{P}_{N N^\prime} \approx 2 \, \mathsf{p}_0 \, \xim^{N+N^\prime} \sqrt{\varrho_N \varrho_{N^\prime}} \cos(\phi_{NN^\prime}) \,.
\end{equation}
The interference phase $\phi_{NN^\prime}$ will be adressed below.

\subsection{Completing the Model: Finding photons in continuous spectra}\label{Sec:Find_Photons}
In order to complete the model, we present in the following a method to determine the probability density for multiphoton events in a laser pulse with a continuous frequency spectrum.
Furthermore, we will discuss how emission processes can be incorporated into the model.

\subsubsection{Probability to find one photon of fixed energy}
We consider the case of a plane-wave fronted pulse, with its electric field being determined by a real function $\ft$. Again, we suppress the spatial dependence in the notation.
Applying Plancharel's Theorem to the calculation of the total energy contained in the pulse
\begin{equation}\label{spectralE}
 \mathcal{E} \sim \int_{-\infty}^\infty dt\, |\ft|^2 \sim \int_0^\infty d\omega |\fw|^2
\end{equation}
where $\fw = \int_{-\infty}^\infty \ft e^{i\omega t} dt$ denotes the Fourier transform of $\ft$, we obtain the spectral energy density $|\fw|^2$. Since we will have to normalize our expressions in order to obtain a probability density, we may drop the prefactors along these first lines.
Introducing photons to our formerly classical calculation, we obtain the photon number density $\frac{1}{\omega}|\fw|^2$ and define 
\begin{equation}\label{One_Photon_Density}
 \pd(\omega) = \frac{1}{N_\pd} \frac{1}{\omega}|\fw|^2 
\end{equation}
as the probability density to find one photon of frequency $\omega$ in the spectrum of the pulse. 
In the following, we  shall refer to $\pd(\omega)$ as the {\it photon finding probability}.
The normalization is achieved with $N_\pd = \int_0^\infty \frac{1}{\omega}|\fw|^2 d\omega $. 
In terms of our model, we understand $\pd_1(\omega) = \pd(\omega)$.

\subsubsection{Finding $N$ photons}
Temporarily excluding emission processes, the probability density to find a combination of $N>1$ photons that sum up to an energy of $\omega$ follows as
\begin{eqnarray}
\pd_N(\omega) &=& \frac{1}{N!} \int d\omega_1 \int d\omega_2 \cdots \int d\omega_N \\
 && \pd(\omega_1) \pd(\omega_2) \cdots \pd(\omega_N)\, \delta\bigg(\omega-\sum_{j=1}^N \omega_j\bigg) \nonumber
\end{eqnarray}
where all frequencies are assumed pairwise different and $\pd(\omega)$ is defined to vanish for non-positive frequencies. 
The expression can be brought into the recursive form
\begin{equation}\label{Photon_Density}
 \pd_N(\omega) = \frac{1}{N} \int_0^{\omega} d\omega^\prime \pd(\omega-\omega^\prime) \pd_{N-1}(\omega^\prime) \,.
\end{equation}
This quantity will be adressed as multiphoton finding probability.

\subsubsection{Negative Photon Numbers -- Emission Processes}
In order to treat emission processes with our model, we draw the analogy to general absorption- and stimulated emission processes. 
The probabilities of both processes are the same (except for degeneracies) and proportional to the photon density at the transition frequency.
Assuming the corresponding pair creation probabilities $\Ppp$ to be well described by $\PppN$, we only need to generalize the multiphoton finding probability $\pd_N$.
This can be achieved by including negatively weighted photons into the summation of $N$ photons with total energy $\omega$ in the form
\begin{equation}
 \pd_N(\omega) = \frac{1}{N} \int_{-\infty}^{\infty} d\omega^\prime \pd(|\omega-\omega^\prime|) \pd_{N-1}(|\omega^\prime|) \,.
\end{equation}
After this amendment, the P-Model is able to account for emission processes.

\subsection{Examples}
This section contains two example cases showing the performance of our model. 
We present an illustrative example of multiphoton processes, followed by an example of emission processes.
The energy spectra are obtained as described in Sec. \ref{Sec:PM:Pre}.
The emission direction of the positron is determined by the azimuthal angle $\phi_{e^+}$ which is measured w.r.t. the laser polarization vector, and by the polar angle $\theta_{e^+}$ being measured w.r.t. the laser propagation direction.

The calculations are performed in a frame of reference in which the gamma quantum frequency and the central laser frequency are of the same order. We note that the parameters for the first example could be achieved in two different experimental scenarios:
Employing a usual Nd:YAG laser with $\omega_c\approx 2.4$eV and peak intensity $\sim 10^{17}$W/cm$^2$, a gamma quantum with $\sim300$ GeV is required, which could be generated by Compton backscattering off an electron beam with ultra-high energy as envisaged by XCELS \cite{XCELS}.
In another scenario, assuming a central frequency of $0.2$ keV as provided by SASE3 at DESY \cite{DESY}, the required intensity is $\sim10^{21}$W/cm$^2$. Various proposals have been put forward of how a few-cycle x-ray laser pulse could be generated (see, e.g., \cite{Saldin2004,Tanaka2015,Prat2015}).
The gamma quantum energy is reduced to 4 GeV in this scenario, which can be achieved with latest laser-plasma accelerators.

\subsubsection{Multiphoton Processes}\label{Sec:Ex_MP}
\begin{figure}[h]
 \includegraphics[width=1.0\columnwidth]{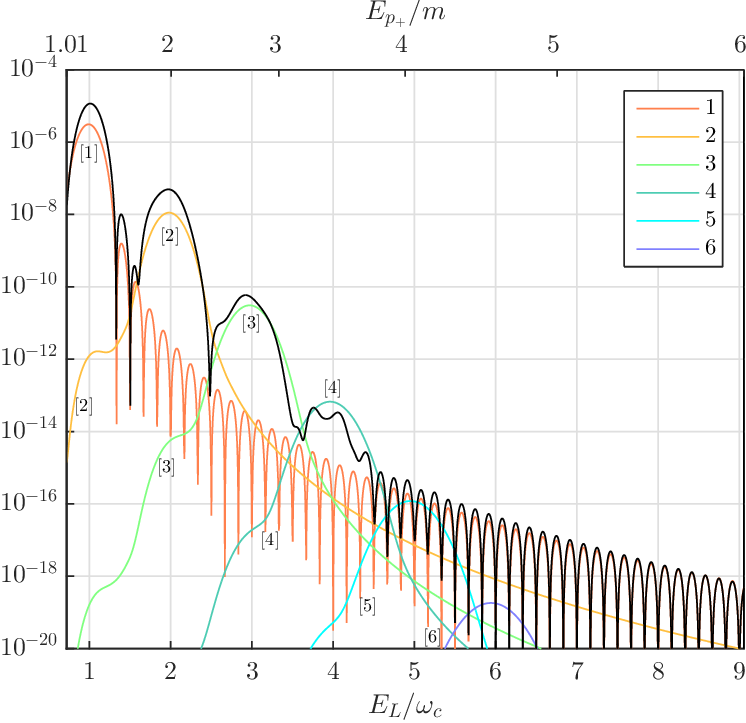}
\caption{\label{fig:Fig1}(Color online) Energy spectrum of positrons $d\mathscr{P}=\frac{d^3 \mathscr{P}}{d E_{p_+} d^2 \Omega_{e^+}} \frac{\partial E_{p_+}}{\partial E_L}(E_L)$ (black line) in units of $1/m$ as a function of the required laser photon energy $E_L$ in units of the central laser frequency $\omega_c$.
Top axis shows the corresponding positron energy.
Colored lines depict model estimates for partial probabilities induced by the absorption of different numbers $N$ of laser photons as indicated in the legend and by the symbols $[N]$.
These positrons with $\phi_{e^+}=\frac{\pi}{4}$, $\theta_{e^+}=0.3\pi$ result from the head-on collision of a laser pulse with $\xi_{\operatorname{max}}=0.1$, $\Nosc=6$, $\omega_c=0.9m$ and $\cep=0$ and a gamma quantum of energy $\omega_\gamma = 3.015m$. 
}
\end{figure}

In Fig. \ref{fig:Fig1} we show a typical energy spectrum obtained from Eq. \eqref{P_Pulse} (black solid line) of positrons as a function of the required laser energy in units of the central laser frequency. 
The spectrum reveals a complicated structure which is dominated by a sequence of broad peaks centered around integer values of $E_L/\omega_c$
(cp. Refs. \cite{Heinzl2010,Krajewska2012BW,Grobe1,Kohlfuerst2014}).
Additional fast oscillations occur especially for higher energies. As an overall impression, the spectrum decays rapidly with increasing energy.

Our model can now be used in order to decode this rich diversity of effects. 
For the current parameters, it is sufficient to regard absorption-only processes 
\footnote{In principle, emission processes also occur for the parameters used in Fig. \ref{fig:Fig1}, but for a given required energy, their probabilities are clearly below the dominant absorption-only process.}.
The model expression for the one-photon channel is obtained from the pulse spectrum and amplitude as described above [see Eqs. \eqref{Model_P_Ord},\eqref{Model_Scaling},\eqref{One_Photon_Density}] and by finally fitting the remaining factor $\mathsf{p}_0$ for the highest energies in the spectrum, with $\mathsf{p}_0\approx 7.0\times10^{-5}$. The resulting estimate is plotted as the red solid line in Fig. \ref{fig:Fig1}, which agrees nicely with the actual pair production probability both at the low- and high-energy part of the spectrum. 
At low energies, a major peak occurs at $E_L \approx \omega_c$, allowing the required laser energy to be provided by one photon of the central frequency of the laser and corresponding to a maximum in the photon finding probability $\pd_1$. 
The model clearly reproduces the approximate shape of the main peak and of the neighboring subpeaks. 
For high energies, the model agrees even quantitatively for a broad range of energies. As a consequence, the fast oscillations are fully explained by the pulse spectrum 
\footnote{We note that the oscillations in the particle spectrum and in the model calculations are offset, when dressing effects are neglected.}.
For both ends of the spectrum, the good agreement leads to the interpretation of the corresponding positrons being produced by a one-photon process.

In order to understand the central part of the spectrum, we determine the model expressions for higher photon numbers, employing the same value for $\mathsf{p}_0$ as for the one-photon process. As can be seen in Fig. \ref{fig:Fig1}, the estimates for the two, three and four photon processes agree well with the shape of the broad peaks which are centered around energies that correspond to the respective multiple of the central frequency. Again, the good agreement supports the identification of the dominant production channel at a given energy.
The oscillating substructures appearing in the higher photon number peaks will be addressed in Sec. \ref{Sec_Num_Res}.

The overall structure of the particle spectrum is determined by the interplay of the perturbative intensity scaling and the fall-off of the photon finding probability, which scales as $(E_L/\omega_c)^{-9}$ for the pulse profile under investigation. 
For the parameters used in Fig. \ref{fig:Fig1}, the one-photon tail clearly exceeds the probabilities of processes involving more than five photons.
Consequently, the number of photons contributing to the dominant pair production channel at a given laser energy $E_L$ cannot simply be deduced from the ratio $E_L/\omega_c$.

\subsubsection{Emission Processes}\label{Sec:Ex_EP}
\begin{figure}[h]
 \includegraphics[width=1\columnwidth]{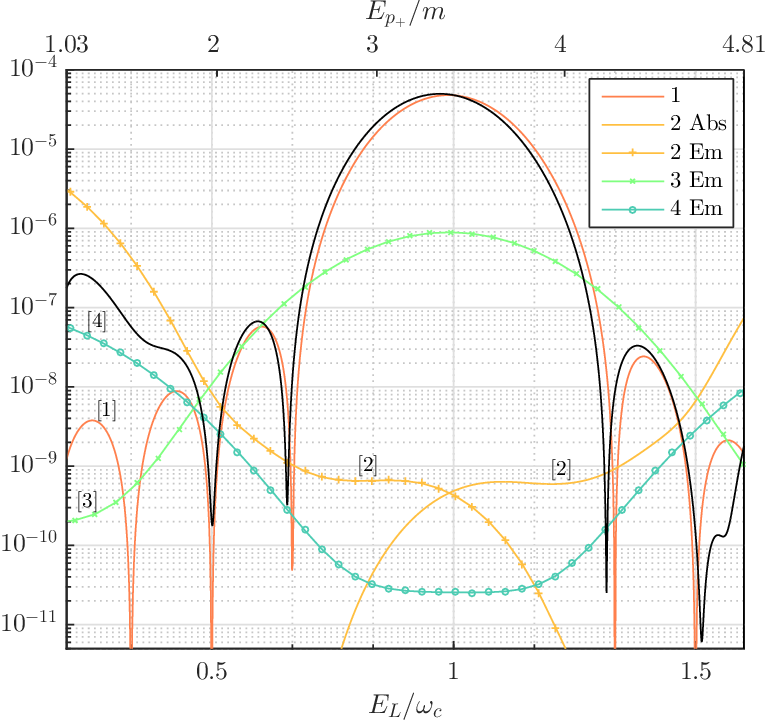}
\caption{\label{fig:EM} (Color online) Positron energy spectrum $d\mathscr{P}$ of Eq. \eqref{dP} (black line) in units of $1/m$ obtained for $\omega_c=3.6m$ and $\xim=0.5$ in order to demonstrate emission processes at low energies. Remaining parameters are identical to Fig. \ref{fig:Fig1}. 
The colored lines depict model estimates for processes involving a total number $N$ of laser photons, where $N$ is indicated in the legend and by the labels $[N]$. The solid lines correspond to absorption-only processes, while the symbol lines show processes where at least one photon is emitted into the laser wave.
The model expressions for the three and four photon absorption-only processes are too small to be seen here.
Subticks indicate laser photon energies $E_L$ which correspond to integer multiples of the laser basis frequency $\omega_b$.
}
\end{figure}

In order to illustrate emission processes, the energy spectrum obtained from a laser pulse with increased amplitude $\xim=0.5$ and frequency $\omega_c=3.6m$ is presented in Fig. \ref{fig:EM} and compared to the model estimates for different production channels. The indicated photon numbers refer to the total number of laser photons involved in a given process. The solid lines correspond to the model calculation for absorption-only processes, while the symbol lines show the model calculation for processes where at least one photon is emitted.
In contrast to the previous figure, the remaining parameter $\mathsf{p}_0$ was chosen in order to obtain good agreement for the major one-photon peak, with $\mathsf{p}_0\approx1.7\times10^{-4}$.

For a production channel involving a given total number of photons, emission processes cause additional broad peaks at lower energies. These peaks are offset by an even number of central frequencies from the main peak in the absorption-only case.  This behavior follows directly from the fact that emitting instead of absorbing one photon changes the resulting energy by two photon energies.

For the parameters used in Fig. \ref{fig:EM}, the pair creation process at small energies is clearly not predominantly caused by a one-photon process, but by a two- (or four-) photon process, where one (or two) of the photons are emitted into the laser wave. 
The enormous laser frequency enables even smallest fractions of the central frequency to produce a pair. The required energy can thus be provided if one photon of the central frequency is absorbed, and the remaining energy is released by emitting one photon with a frequency slightly less than $\omega_c$.

We note that with decreasing values of $\xim$, the relative importance of processes with higher total photon numbers decreases, and the particle energy spectrum approaches the shape of the one-photon process. The characteristic zeros of the one-photon process [see Fig. \ref{fig:EM}] at energies corresponding to integer multiples of the laser basis frequency will be addressed below.

Concluding the section, we have shown that our approach allows to understand the position of the main peaks in the spectrum, as well as their approximate shapes. In the following, we will investigate the finestructure further and detect CEP and interference effects.

\section{CEP and Interference Effects}
\label{Sec:Int}
In this section, we investigate how the carrier-envelope phase enters the pair creation probability in the multiphoton regime. The first subsection contains an analytical approach for a general pulse shape, while the second subsection contains numerical examples and shows how our model can be used to identify interference effects.

\subsection{CEP Effects -- Analytic Approach}
In the first part, we establish a connection between the phase shifts $\delta_j$ of Eq. \eqref{Multichrom_Field} and the spectral phase of a finite pulse.
In the second part, we will investigate how the pulse spectrum and the spectral phase are affected by the CEP.
The third part applies these findings to the pair creation process.

\subsubsection{Continuous generalization of the phase shifts $\delta_j$}
Let us regard a plane-wave fronted pulse which is determined by its vector potential $A(t)$.
We regard the Fourier decomposition in the form 
\begin{equation}
 A(t) = \frac{1}{\pi} \int_0^\infty |\hat{A}(\omega)| \cos(\omega t - \phi_\omega) d\omega
\end{equation}
where the spectral phase $\phi_\omega = \arg \hat{A}(\omega)$ has been introduced. A comparison with the multichromatic field Eq. \eqref{Multichrom_Field} shows that the phase shift $\delta_j$ can be associated with the spectral phase $\phi_\omega$ of the corresponding frequency mode. The latter can equally be expressed by the spectral phase of the electric field $E(t)$, such that we obtain
\begin{equation}\label{cont_gen_phaseshift}
 \delta_j \, \widehat{=} \, \arg \hat{A}(\omega) = \arg \hat{E}(\omega) - \pi/2
\end{equation}
which determines the relevant components in the phase of Eq. \eqref{IntTerms}.

\subsubsection{CEP signatures in pulse spectrum}
\label{Sec:CEP_in_Pulse_Spectrum}
Let us assume, for symmetry reasons, the electric field of a plane-wave fronted pulse to be given by
\begin{equation}
 f(t) = f_{\text{env}}(t) \cos(\omega_c t + \cep)
\end{equation}
with an arbitrary envelope $f_{\text{env}}(t)$ that does not depend on the CEP $\cep$.
The Fourier transform has the form
\begin{equation}\label{CEP_in_FT}
 \hat{f}(\omega) \sim \hat{f}_{\text{env}} (\omega+\omega_c) e^{i\cep} + \hat{f}_{\text{env}} (\omega-\omega_c) e^{-i\cep}
\end{equation}
revealing the structure of the pulse spectrum and its explicit dependence on the CEP. 

The CEP dependence becomes particularly straightforward, if the first term can be neglected, which requires the following conditions: 
(i) the spectral width of the envelope needs to be small, as for not-too-short pulses;
(ii) we restrict ourselves to the most dominant frequency components $\omega$ which are close to the central frequency $\omega_c$.
Under these assumptions, the resulting photon density does not depend on the CEP, while the spectral phase reveals a linear dependence on the CEP.

If one of these conditions is not fulfilled, such as in the high-energy part of the spectrum, the photon density and the spectral phase possess a more complicated CEP-dependence.

\subsubsection{CEP effects in energy spectra of Breit-Wheeler particles}
\label{Sec:CEP_Analytical}
Combining the previous findings, we can understand the influence of the CEP on the pair creation process as seen in the corresponding energy spectra of the particles.

For the following discussion, the pulse shape is not specified but assumed to be in accordance with condition (i).
The simplified CEP dependence on the pulse spectrum therefore applies for those processes which are mainly induced by photons with frequencies around $\omega_c$.
For a typical particle energy spectrum, this condition holds in the interval comprising several multiphoton peaks, starting at energies around $\omega_c$ (when emission processes are negligble), and extending to the energy at which the one-photon process becomes noticeable again.

In this inner part of the particle spectrum, the relevant multiphoton finding probabilities $\PphN$ are insensitive to the CEP. 
On the other hand, the CEP dependence of the spectral phase is conveyed to the phase shifts, which determine the interference phases.
Let us further assume the envelope function to be symmetric under time-reversal, such that the complex phase of its Fourier transform does not obtain a physically relevant continuous frequency dependence.
Thus, for interference terms between (absorption-only) processes with $\Nphotons$ and $\Nphotons^\prime$ photons, the interference phase obtains a $(\Nphotons-\Nphotons^\prime)\chi$ dependence [cp. Eq. \eqref{Int_N_Photon}]. This phase can lead to pronounced interference effects, whose visibility is determined by the probabilities of the underlying channels
\footnote{For self-interference terms, this CEP-dependence disappears. Concerning the $\cep$-dependence of emission processes, recall Eq. \eqref{AmplitudePhase}.}.
We note that our pulse [cp. Eq. \eqref{our_pulse}] as employed in the numerical examples fulfills both criteria.

In the outer part of the particle spectrum, both the multiphoton finding probabilities and the interference phases have more complicated CEP dependences.

\subsection{Numerical Results}\label{Sec_Num_Res}
\begin{figure*}
 \includegraphics[width=1.0\textwidth]{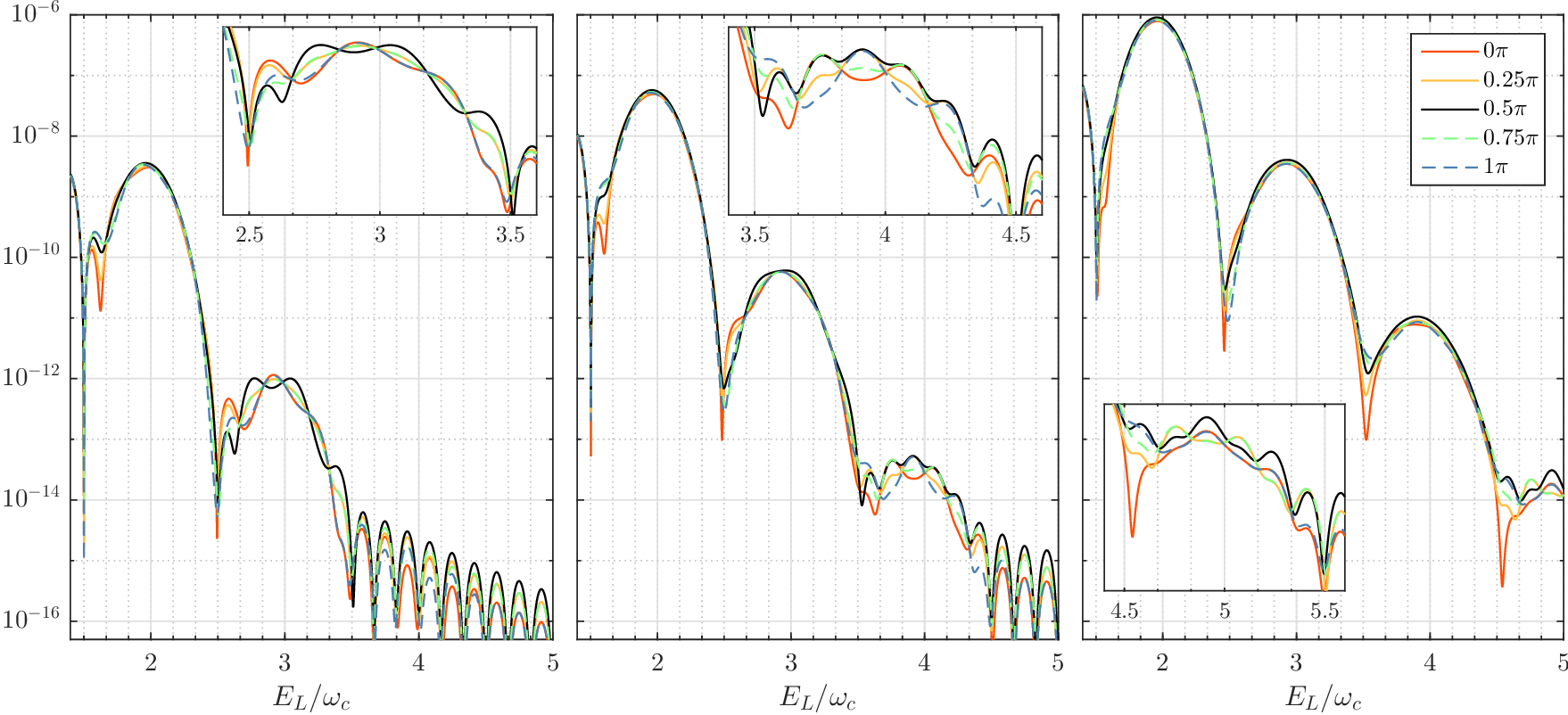}
\caption{\label{fig:CEP} (Color online) Positron energy spectra $d\mathscr{P}$ in units of $1/m$ for various values of the carrier-envelope phase as indicated in the legend and for $\xim=0.05$ (left), $\xim=0.1$ (center) and $\xim=0.2$ (right). 
Remaining parameters are the same as in Fig. \ref{fig:Fig1}.
The inlets show the respective last multiphoton peak before the one-photon tail begins.
Subticks indicate laser photon energies $E_L$ which correspond to integer multiples of the laser basis frequency $\omega_b$.}
\end{figure*}

In this section, we will demonstrate and explain CEP effects occuring in the particle energy spectra for our specific choice of the pulse shape [cp. Eq. \eqref{our_pulse}]. 
To this end, we present numerically computed particle spectra in Fig. \ref{fig:CEP} for various values of the CEP $\cep$ while the maximum pulse amplitude $\xim$ is kept constant. This normalization simplifies the analysis, since, in the spirit of the P-Model, the probabilities of the dominant pair production channels are kept constant. 
Note, however, that the corresponding pulse energy is not constant but depends on $\cep$.
The pulse amplitude is varied in the panels with $\xim=0.05$ (left), $\xim=0.1$ (center) and $\xim=0.2$ (right). 
Each increment of $\xim$ facilitates one more multiphoton peak before the one-photon tail begins. These last peaks are associated with photon numbers $\tilde{N}=3,4$ and $5$, respectively.

For all depicted values of $\xim$, the CEP is found to modify the particle spectra locally, while their general structure as determined by the multiphoton peaks and the one-photon tail is preserved.
Between neighboring multiphoton peaks, quantitatively strong CEP effects are found that can change the pair creation probability by more than one order of magnitude.
For a given value of $\cep$, especially the peak associated with $\tilde{N}$ (as depicted in the inlets) is modulated by small  oscillations with a scale given by the laser basis frequency $\omega_b$. These modulations reveal a strong dependence on $\cep$. 
The one-photon tail exhibits a strong dependence on $\cep$ as well. 

In order to understand these effects, we follow the previously developed analytical approach. With $\Nosc=6$, the pulse shape is chosen in accordance with condition (i) from Sec. \ref{Sec:CEP_in_Pulse_Spectrum}.
Condition (ii) is fulfilled for all processes which are not affected by the high-energy CEP-dependence of the photon finding probability $\varrho_1$. The latter begins at $E_L \approx 3\omega_c$ (not shown). Determined by the strength of the three- and four-photon peaks, the inner part of the particle spectra as introduced in Sec. \ref{Sec:CEP_Analytical} extends to $E_L\approx 3.5\omega_c$ for $\xim=0.05$ (left panel), and to $E_L\approx4.5\omega_c$ in the center and also the right panel.

In the inner parts, the distinct CEP effects can be attributed to interference effects. 
At energies in-between multiphoton peaks, neighboring photon-number channels are of the same order and can induce strong interference terms, with an interference phase being proportional to $\cep$. This effect is the main origin of the CEP-dependences which can be seen in Fig. \ref{fig:CEP} at $E_L \gtrsim 1.5\omega_c$ in all panels, at $E_L \gtrsim 2.5\omega_c$ in the center and right panel and at $E_L\gtrsim 3.5\omega_c$ in the right panel.
The CEP dependence was analyzed by means of a Fourier decomposition of $d\mathscr{P}(\cep)$ for fixed values of $E_L$.
We further note that the CEP effects can become more involved when interferences between other photon number channels contribute noticeably to the pair creation probability.

The multiphoton peaks are also subject to interference effects. 
The difference between the probabilities of the interfering channels generally decreases as the number of photons is increased (cp. Fig. \ref{fig:Fig1}) and gives rise to the most prominent interference effects in the peak associated with $\tilde{N}$.
Conversely, for a given multiphoton peak, this difference grows with $\xim$ and weakens the interference effects.
The small spectral oscillations disappear simultaneously, leading to the conclusion that they are mainly caused by interferences
\footnote{
In the inner part, all CEP effects visible in Fig. \ref{fig:CEP} allow an interpretation in terms of ordinary interferences. 
We have not found signatures of self-interferences. They could be expected to be prominent on top of those multiphoton peaks which have no competing processes, such that ordinary interferences are negligble. Still, these peaks are almost invariant under variations of $\cep$. This finding is in good agreement with the analytical approach predicting a vanishing $\cep$-dependence of the corresponding interference phase. 
}. 

Let us deepen the discussion by analyzing the peak associated with $\tilde{N}$. Here, interferences can be expected to happen predominantly between the channels involving $\tilde{N}$ photons and the one-photon channel. The corresponding interference phases have a periodicity of $2\pi/(\tilde{N}-1)$ in $\cep$. 
These interferences are the main reason for the $\cep$-dependence of the spectral oscillations as depicted in the inlets of Fig. \ref{fig:CEP}. In the right panel, the moderate $\cep$-dependence of the one-photon channel additionally enhances the pair creation probability for $\cep=\pi/2$ as compared to $\cep=0,\pi$. 
For this reason, the inner part of this spectrum is defined to end at $E_L\approx 4.5\omega_c$.

In order to complete the picture, we regard the shape of the corresponding spectral oscillations.
In the left panel, they almost vanish for $\cep=\pi/4$ and $\cep=3\pi/4$. While their original periodicity is $\omega_b/2$, the remaining oscillatory structure has a periodicity of $\omega_b$. In the center panel, a similar behavior is found, with smallest deviations from the original shape of the multiphoton peak for $\cep=\pi/2$. 
In the right panel, the corresponding behavior would appear for $\cep=\pi/8$ (not shown).

Considering additional properties of the pulse shape under investigation, these observations can be explained.
The pulse spectrum contains strong signatures of the finite temporal length of the pulse.
The Fourier transform of the characteristic function is of the form $\operatorname{sinc}(\pi\omega/\omega_b)$.
As a consequence, the pulse spectrum contains zeros at integer values of $E_L/\omega_b$ (except for energies in the main central peak, cp. Fig. \ref{fig:EM}), where the spectral phase jumps by $\pi$. 
The particle spectrum and in particular the spectral oscillations are affected by both effects:
While the characteristic zeros determine the energy dependence of the one-photon process, the phase jumps lead to a discrete energy dependence of the interference phases (see Appendix for further details).

Accordingly, the shape of the spectral oscillations is determined by the one-photon-process and additionally modulated by the energy-dependent interference phase. The latter is given by $(\tilde{N}-1)\cep+j\pi$ where the integer $j$ is increased when $E_L$ passes integer multiples of $\omega_b$. When the interference terms vanish due to a specific choice of $\cep$, the remaining oscillatory structure results from the (incoherent) addition of the one-photon process on top of the multiphoton peak.

The one-photon tail lies in the outer part and reveals further CEP effects which are mainly determined by the high-energy CEP-dependence of $\varrho_1$. Additionally, they can be caused by interferences involving e.g. the (weak) channel with $\tilde{N}+1$ photons.

Before proceeding to the conclusion, we would like to draw a brief comparison between our model and a related model which was presented in \cite{Nousch2012}.
Both approaches decompose the pair creation process into contributions of different photon numbers. The model in \cite{Nousch2012} applies exactly determined pair production cross sections (integrated over the emission angles) in monochromatic fields which are, as an approximation, convoluted with the pulse profile. Contrary to that, in our model an approximation to the pair production probability is used while the emphasize is laid on the spectral properties of the laser pulse. Hence, both models possess a conceptually different structure.

\section{Conclusion}\label{Sec:Conclusion}
In this paper, we have investigated the strong-field Breit-Wheeler process in short laser pulses with intermediate intensities $\xi\le1$. 
Employing detailed $S$ matrix calculations, the energy spectra of emitted particles have been investigated, which exhibit a rich structure.
Generalizing concepts from a bichromatic to a multichromatic laser field and regarding the finite laser pulse as a limiting case, 
our approach is based on the spectral properties of the laser pulse. This approach enabled us to understand the structure of the energy spectra as well as the effects of the laser carrier-envelope phase. 

The carrier-envelope phase was found to have a two-fold impact on the particle spectra. First, it influences the probabilities of multiphoton process. Second, it directly affects the interference phase between different pair production channels. The latter effect clearly reflects the analogy to the role of the relative phase shift in a bichromatic field of orthogonal polarization (see, e.g., \cite{Augustin2013,Jansen2015}).
The combination of both effects leads to distinct signatures of multiphoton interferences in the particle spectra.

Our approach has led to an intuitive model based on the probabilities of multiphoton processes. 
The model has supported the analysis of the particle spectra with quantitative estimates for the magnitudes of different production channels, and has additionally allowed to detect combined emission-absorption processes.
It can easily be extended to include high photon numbers and can further be applied to gain insights into other multiphoton QED processes in intense laser pulses as well.

\section*{Acknowledgement}
Fruitful discussions with M. Dellweg, J. Z. Kami\'{n}ski, S. Meuren and A. B. Voitkiv are gratefully acknowledged.
This study has been performed within project B11 of SFB-TR18 funded by the German Research Foundation (DFG).

\section*{Appendix}
In this appendix, we shall give further details on the dependence of the amplitude phase, as briefly described in Sec. \ref{Sec_Num_Res}. 
First, we address the phase jumps in the spectrum of our particular laser pulse [cp. Eq. \eqref{our_pulse}].

Let us regard the interference terms in Eq. \eqref{IntTerms}. 
The phases $\varphi(\mn)$ are determined by the spectral phases $\phi_\omega$ of the participating photons [cp. Eq. \eqref{cont_gen_phaseshift}].
Let us further regard processes in the inner part of the particle's energy spectrum as defined in Sec. \ref{Sec:CEP_Analytical}, 
such that the spectral phase of a relevant photon with frequency $\omega\equiv\omega_j$ is given by 
\begin{equation}\label{sign_change_phi_omega}
 \phi_{\omega_j} = -\chi + \ell_j \pi + \text{const}\,.
\end{equation}
The integer $\ell_j$ is zero if $\omega_j$ is equal (or close to) $\omega_c$. It is increased by one whenever $\omega_j$ passes integer multiples of $\omega_b$ outside the main spectral peak, which is located within the interval $(\omega_c-2\omega_b,\omega_c+2\omega_b)$, cp. Figs. \ref{fig:EM} and \ref{fig:Fig1}. 
The constant in Eq. \eqref{sign_change_phi_omega} does neither depend on the frequency $\omega_j$ nor on $\cep$.
For a given photon combination $\mn$, we introduce the total number of sign-changes $\ell_{\mn} = \sum_{j=1}^N \ell_j$.

Now we can decompose the full interference term in Eq. \eqref{IntTerms} into contributions from certain photon combinations $(\mn,\mnp)$ being sorted by the number of sign-changes $\Delta \ell_{\mn,\mnp} = \ell_{\mn}-\ell_{\mnp}$.
At this point, we can also account for sign-changes induced by $\sigma_{\mn,\mnp}$ [cp. Eq. \eqref{IntTerms}].
Since all these sign-changes affect the global sign of a given interference process between two photon combinations $\mn$ and $\mnp$, we end up with the full interference term $\mathcal{P}_{NN^\prime}$ being divided in two addends: One with a constructive interference, the other one with destructive interference. 
With $\Delta \ell_{\mn,\mnp}$ being independent of $\cep$ in the inner part of the spectrum, and assuming the same for $\sigma_{\mn,\mnp}$, both interference terms have an effective interference phase of $(N-N^\prime)\chi$.

Second, for interference processes being affected by the high-energy CEP-dependence (see e.g. the right panel of Fig. \ref{fig:CEP}), the interference phase may be more complicated than $(N-N^\prime)\cep$. In fact, when the first term in Eq. \eqref{CEP_in_FT} has to be taken into account, the resulting spectral phase deviates from the linear dependence on $\cep$. Still, these deviations are found to have only small impact on the appearance of the corresponding $\cos$-term in our numerical computations.

\end{document}